\documentstyle[twoside,fleqn,espcrc2,epsfig]{article} 

\newcommand{\AmS}{{\protect\the\textfont2 
  A\kern-.1667em\lower.5ex\hbox{M}\kern-.125emS}}

\renewcommand{\Im}{\mathop{\mathrm{Im}}} 
\renewcommand{\Re}{\mathop{\mathrm{Re}}}

\title{Kondo effect in mesoscopic systems}

\begin{document} 
 
\author{O. \'Ujs\'aghy \address{
Institute of Physics and Research Group of Hungarian Academy 
of Sciences, Technical University of Budapest, 
H-1521 Budapest, Hungary}\thanks{OTKA postdoctoral Fellow D32819},  
G. Zar\'and$~^{\mathrm{a}}$, and 
A. Zawadowski$~^{\mathrm{a}}$\address{ 
Research Institute for Solid State Physics, POB 49, H-1525 
Budapest, Hungary}}

\begin{abstract} 
The Kondo effect may develop  in those cases where there are non-commuting 
operators describing the interaction between the conduction electrons 
and impurities or defects  with internal degrees of freedom. 
This interaction may involve spin 
or orbital variables. There are cases, where the conduction electrons 
have  conserved quantum numbers which do not appear in the coupling.
An example is 
where  the Kondo effect involves orbital degrees 
of freedom, and the interaction 
is independent of the real spins
of the electrons, which is conserved and leads to the 
two-channel Kondo (2CK)  problem. The low temperature 
behavior is very different for the one 
and two-channel cases, as in the first case  a Fermi 
liquid is formed while in the second 
one strong deviations  appear and a non-Fermi liquid state is realized. 
Mesoscopic samples provide 
a unique possibility to study 
a few or even a single Kondo impurities.
In this paper we first review how the 
original spin Kondo  problem is affected by  surface anisotropy
and the fluctuations  of the density of 
states in point contacts. We discuss 
the physics of a single magnetic impurity 
on the surface of a sample.
The orbital 2CK effect 
due to dynamical defects  is considered. The nature of these defects is 
not known but they are excellent candidates to describe the 
zero-bias anomalies with non-Fermi liquid character in point contacts, 
and the dephasing time and transport in short wires. There are two main 
concerns with this interpretation: Firstly, the tunneling centers 
formed by heavy impurities may 
produce too small Kondo temperature, and secondly, 
the splittings seen in the experiments are much smaller than expected
from this model.
It would be therefore extremely important to identify the microstructure 
of these two level systems and find new realizations for them.
\end{abstract} 
\maketitle

\section{Introduction} 
\label{S1} 
 
Since the discovery  of the anomalous low temperature resistivity increase  
exhibited by some metallic samples   \cite{Haas}   
these anomalies  attracted considerable interest. The first  
theoretical work to explain them  was due to Kondo,  
who demonstrated that the scattering rate of electrons in metals by magnetic  
impurities has an anomalous third order contribution,  which increases 
  logarithmically as the temperature is reduced  
and leads to the break-down of perturbation theory 
\cite{Kondo}. Since then this phenomenon is known as the  
Kondo effect.  
Following Kondo's original work a lot of theoretical effort 
has been devoted to understand this phenomenon in detail. 
Wilson's  numerical renormalization group  
to treat the strong coupling limit \cite{Wilson} 
and Nozi\`eres' Fermi liquid theory \cite{Nozieres} 
turned out to be  
the most important  milestones  in this development.  
 
Recently,  the number of papers related to the Kondo effect showed 
a significant increase with 
broader and broader applications of the model. 
Various dilute and dense $U$ and $Ce$ based 
metallic alloys have been suggested as Kondo systems with 
both magnetic and orbital features \cite{Cox}. In   
these systems at low temperature very strong correlations 
build up, hence they became known as strongly correlated 
systems. Other new developments were in the direction of the observation 
of  Kondo   effect in  mesoscopic systems such as thin layers and 
point contacts, 
 and also artificial  mesoscopic atoms (quantum dots).  
In these latter nanofabricated devices   
the d-level of the magnetic impurity in the metal 
is mimicked by degenerate states of a  quantum dot,  
which is coupled to metallic or semiconducting leads. 
 
Nanotechnology  itself is a very  fast developing field, 
which opens up new perspectives and offers new possibilities 
to study magnetic impurities and strongly correlated systems.  
Of course, its extensive overview or a discussion of  the physics of  
 nanofabricated  artificial atoms is  out  of the scope of  
our review. Here we only focus on the study of magnetic and dynamical  
impurities in  mesoscopic systems.

The Kondo effect, in general, originates from the 
scattering of  conduction electrons by a localized object  
(magnetic or substitutional impurity  or some topological  
defect) with some internal degrees of freedom  (e.g. spin, two close  
atomic positions, dislocation kink). 
The typical Hamiltonian of Kondo-like problems is 
\begin{eqnarray} 
H&=&\sum_{k\mu} \epsilon_k c^{\dagger}_{k\mu}c_{k\mu}  + 
\sum_{\alpha}\epsilon_\alpha b^{\dagger}_{\alpha}b_{\alpha}   
\nonumber \\ 
& + & \sum_{k,k'}\sum_{\mu\nu\alpha\beta} V_{\mu\nu}^{\alpha\beta} 
c^{\dagger}_{k\mu} 
c_{k'\nu} b^{\dagger}_{\alpha}b_{\beta}  
\end{eqnarray} 
where $\epsilon_k$ is the electron kinetic energy with momentum $k$,  
 $c^{\dagger}_{k\mu}$  creates an 
electron spherical wave with radial momentum 
$k$ and internal quantum numbers $\mu$  
and $b^{\dagger}_{\alpha}$ 
creates a heavy object  with quantum number $\alpha$ ($\alpha$ 
being the spin, the position, or a crystal field  label of the impurity).   
Note that the internal indices $\mu$, $\nu$ of the conduction electrons 
may also represent magnetic spin or  
orbital indices or a combination of them as well. 
 $V_{\mu\nu}^{\alpha\beta}$ denotes the interaction potential  
and a band cutoff $D$ (usually of the order of Fermi energy)  
is applied for the conduction electrons. 
 
\begin{figure} 
\parindent=2.in 
\indent{ 
\epsfxsize=2in  
\epsffile{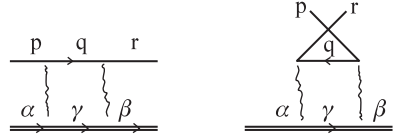}} 
\parindent=0.5in 
\caption{The two leading diagrams for the scattering of a light 
electron on a 
heavy object.  The light electron is represented by the light line and 
the heavy 
object by the double line.   The wavy lines indicate the interaction 
matrix element 
$V$.  The quantum numbers characterizing the particles are also 
shown.    
} 
\label{fig1p2} 
\end{figure}

The first corrections  to the electron-impurity scattering matrix 
are given by the two time-ordered diagrams shown  in Fig. ~\ref{fig1p2}. 
The direction of time corresponds to the direction of the lines on the 
heavy objects.  Assuming an  
interaction independent of $k,k'$, the scattering amplitude 
for an incoming 
electron with energy $\omega$ is 
\begin{equation} 
{V^{(2)}}_{\mu\nu}^{\alpha\beta}(\omega) = \sum_{\rho\gamma} 
[V_{\mu\rho}^{\alpha\gamma} 
V_{\rho\nu}^{\gamma\beta} - 
V_{\rho\mu}^{\alpha\gamma}V_{\nu\rho}^{\gamma\beta}]\ln({D\over\omega}), 
\label{eq:leading}
\end{equation} 
where the quantum numbers of the internal lines are summed over and the 
negative sign arises from the fermion anticommutation relations (note the 
crossed lines in the 
second diagram).  
The logarithm above was first identified by Kondo.  
The divergence of this term as $\omega \to 0$ reflects the break down
of perturbation theory.  
 
Different impurity models  can be classified by the value of the 
``commutator'' of Eq. (\ref{eq:leading}):
{\it (i)} For the {\em  commutative model} 
$[V_{\mu\rho}^{\alpha\gamma} 
V_{\rho\nu}^{\gamma\beta} - 
V_{\rho\mu}^{\alpha\gamma}V_{\nu\rho}^{\gamma\beta}] = 0$, and  
no Kondo logarithms appear.  An example for this is  
the dissipative tunneling system where the interaction  
between the heavy object and conduction electrons is diagonal in 
the internal indices \cite{E}.   
{\it (ii)} For a  {\em non-commutative model} 
$[V_{\mu\rho}^{\alpha\gamma} 
V_{\rho\nu}^{\gamma\beta} - 
V_{\rho\mu}^{\alpha\gamma}V_{\nu\rho}^{\gamma\beta}] \ne 0 $, and 
logarithmic terms appear 
in the scattering matrix. In the following we only concentrate to the  
second case.

In many cases, the general form of the interaction Hamiltonian  
can be simplified by introducing appropriate variables and truncating the 
Hilbert space to 
\begin{equation} 
\nonumber 
  H_{int}=\sum\limits_{\scriptstyle i=x,y,z \atop 
  \scriptstyle \alpha,\beta=\pm, s=1...n} V^i 
  b_{\alpha}^\dagger \sigma^i_{\alpha\beta} b_{\beta} 
  c_{k\mu s}^\dagger \sigma^i_{\mu\nu} c_{k'\nu s} \;,
\end{equation} 
where $\sigma^i$'s are the Pauli operators, $V^i$'s are 
anisotropic couplings and the conduction electron may have an 
additional channel index $s$ which is conserved and does not occur in the 
coupling itself. 
As it has been pointed 
out by Nozi\`eres and Blandin \cite{Blandin} its mere existence can change  
the nature of the low temperature behavior drastically. According to 
the possible different values of that channel indices we speak about n-channel 
Kondo problem ($n=1,2,\dots$). 
 
In the original spin Kondo problem $V^i=J/2$ 
where $J$ is the exchange 
coupling between the localized spin described by the spin indices 
$\alpha,\beta$ and the conduction electron spins labeled by 
$\mu,\nu$. $b_{\alpha}^\dagger$ creates the localized electron state 
with spin $S$, for simplicity $S=1/2$. In that model there is no 
additional channel, thus $n=1$. In this case the localized $S=1/2$  
 spin is screened by the ``compensation cloud'' \cite{felho} 
of the 
electrons and finally a singlet is formed. The binding energy  
is proportional to the Kondo temperature $T_K$ which for the  
isotropical case can be written as  
\begin{equation} 
T_K = D\; (2\varrho_0 J)^{1/2} {\rm exp} (-1/2\varrho_0 J)\;, 
\label{eq:T_K} 
\end{equation} 
with $\varrho_0$ the density of states of the conduction electrons at the  
Fermi level for one spin direction.    
At $T\ll T_K$ thermally excited conduction electrons cannot break the 
singlet, which acts  
as a rigid potential scatterer and Nozi\`eres' Fermi liquid  
theory holds \cite{Nozieres}.  
 
The actual size of the compensation cloud is given by the Kondo
coherence length 
\begin{equation} 
\xi_K \sim {v_F\over T_K}\;, 
\end{equation} 
with $v_F$ the Fermi velocity. For $T_K\sim 1K$ it can  easily exceed 
$1\mu m$, which is in the range of the size of a mesoscopic sample or
even can be much larger. This  
observation triggered the experimental study  of the dilute Kondo alloys  
in thin films, wires and point contacts (see Sec.~\ref{S2A} and \ref{S2B}).  
 
Contrary to the $n=1$ case, for the two channel Kondo model ($n=2$)   
the ground state is not a singlet because two electrons with different 
channel indices compete to screen the impurity spin. The complexity of the  
ground state is reflected in a residual entropy of  
$S=k_B \;\ln\sqrt{2}$, \cite{Bethe}  
which shows that the impurity is only partially screened even at $T=0$ 
and that  small energy excitations are not frozen out even there.   
As a result,
the linear specific heat coefficient $C(T)/T$ and the impurity 
susceptibility were 
found to diverge logarithmically.  
A further surprising result arose from the conformal field theory approach:  
Affleck and Ludwig \cite{C} showed that (i) the impurity contribution  
to the resistivity shows a $\sqrt{\max\{\omega,T\}}$ singularity (as opposed  
to the $\sim \omega^2, T^2$ Fermi liquid behavior) (ii) that the
amplitude of scattering to a one-electron state 
vanishes at  $T=\omega = 0$, and an incoming electron ``evaporates'' to  
infinitely many electron-hole excitations once it hits the impurity  
\cite{B,C} (see Table~\ref{table}).  
 
It must be emphasized  that for $n=2$ an arbitrarily  small  
splitting of the impurity states (produced by magnetic field  
or a strain field for dynamical defects)  
provides a cutoff for the non-Fermi liquid behavior and  
ultimately leads to a crossover to a Fermi liquid state.  
 
\section{Spin Kondo effect in mesoscopic devices} 
\label{S2} 
 
\subsection{Size dependence in films and wires} 
\label{S2A} 
 
\begin{figure} 
\epsfig{figure=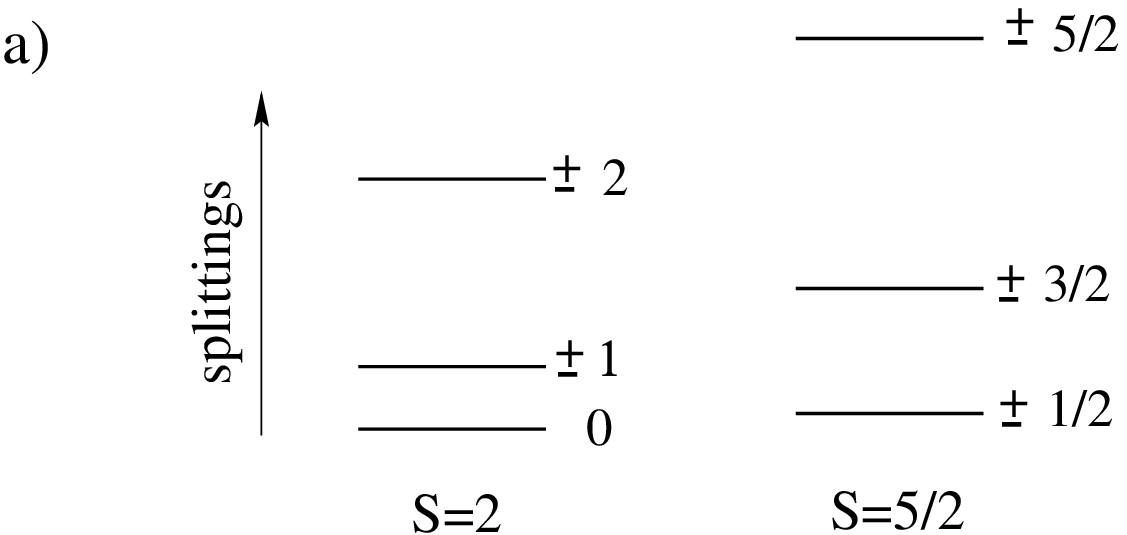,height=3truecm} 
\vskip0.5truecm 
\epsfig{figure=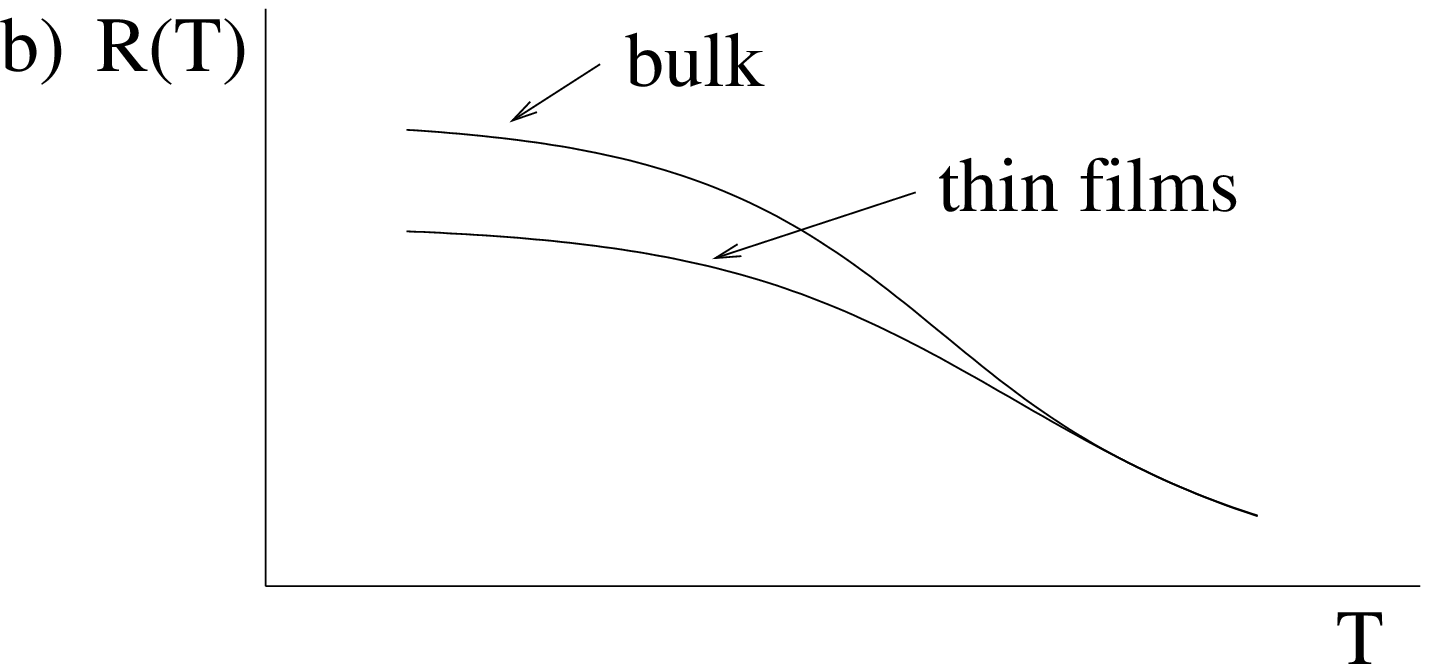,height=3truecm} 
\epsfig{figure=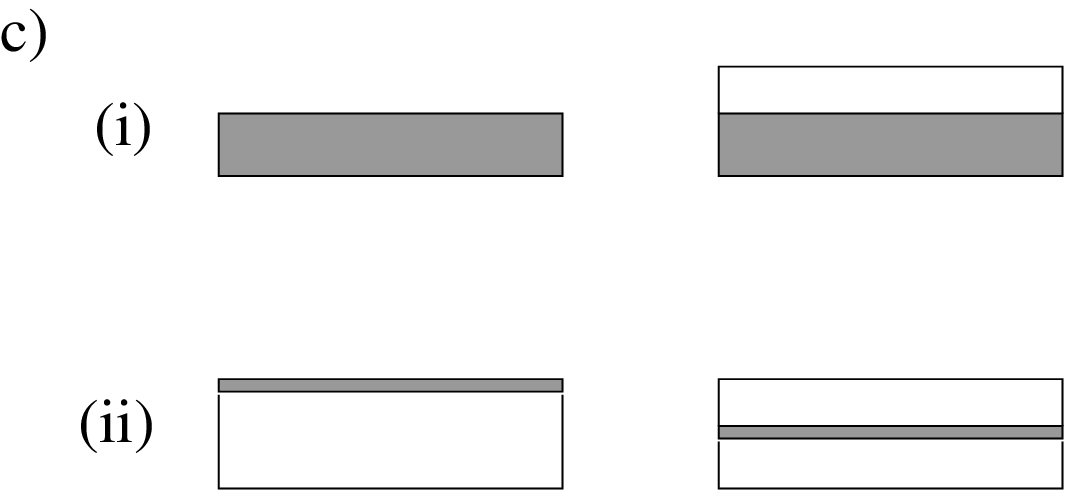,height=3truecm} 
    \caption{Surface anisotropy: (a) the level splitting described by 
    Eq.~(\ref{aniz}) for integer ($S=2$) and half-integer ($S=5/2$) 
    spins. 
(b) Schematic plot of the change in the Kondo resistivity in thin 
    films. (c) The setups of experiments where the shadowed areas 
    represent the dilute magnetic alloy and the clean ones the pure 
    metal. 
The Kondo amplitude increases as (i) the alloys are covered by pure 
    metal \cite{prox,Blachly} (ii) the position of thin layer of the 
    alloy is moving away from the surface \cite{Giordanolayer}.} 
\label{figaniz} 
\end{figure}

In the last decade, 
many experiments \cite{BG,exp1,exp2,Roth}  
were performed on thin films and narrow wires of dilute magnetic 
alloys in search of the  Kondo compensation cloud.  
In these experiments (see Fig.~\ref{figaniz}) no essential 
change in the Kondo temperature was observed, 
however, in most of them  \cite{BG,exp1} a suppression  
of the Kondo resistivity amplitude  
was observed for small sample sizes. 
Covering a thin layer of magnetic alloys by 
another pure metal layer, a partial recovery  
of the Kondo signal  was found  
\cite{prox,Blachly} which was smaller for more disordered overlayers  
\cite{proxdis}.  
\noindent 
\noindent  
The first natural explanation concerning the 
compensation cloud \cite{Bergmann} was ruled out both theoretically  
\cite{Affleck,Zarand} and experimentally \cite{Blachly} as  
the Kondo singlet is formed whenever the level spacing 
is small compared to the Kondo temperature: $\delta\epsilon < T_K$. 
The effect of local density of states (LDOS) fluctuations  
close to the surface (discussed in the next subsection) is also
probably relatively small for the investigated alloys \cite{ZarandUdvardi}.  
 
Two theories have been developed, that seem to explain the two 
limiting cases in the experiments: 
 The first, based on weak localization \cite{Phillips}, 
may be valid in disordered samples 
\cite{BG,prox}, where the smallest system  size is large compared to 
the elastic mean free path and the Kondo anomaly depends 
on the level of disorder. The other explanation,  
the theory of  spin-orbit-induced surface 
anisotro\-py \cite{UZ} explains all the experiments performed in 
ballistic samples (i.e., when the size of the sample is in the 
ballistic region). This  surface anisotropy is 
developed in samples with strong spin-orbit interaction  
on the non-magnetic host atoms \cite{UZ}.  
In this case electrons can mediate information 
about the {\em geometry} of the sample resulting in an anisotropy for the 
impurity spin nearby the surfaces, but only in those cases where the
angular momenta of the localized orbital $l\neq 0$ (e.g. $l=2$). According to  
Ref.~\cite{UZ} this anisotropy is non-oscillating in the leading order 
and inversely proportional to the  
distance $d$ measured from the surface.   
For flat  surfaces it is described by the Hamiltonian 
\begin{equation} 
\label{aniz} 
H_a=K_d ({\bf n S})^2 
\end{equation} 
where ${\bf n}$ is the normal vector of the  surface 
and ${\bf S}$ is the spin operator of the impurity. 
The anisotropy factor $K_d$  is positive and  
is in the range of  
$\frac{0.01}{(\textrm{d}/\textrm{\AA})}\,eV < K_d < \frac{1}{(\textrm{d}/ 
\textrm{\AA})}\,eV$ \cite{UZ}. 
An elegant extension of these calculations to general geometries was 
performed  by Fomin and coworkers \cite{Fomin} who investigated  the 
dependence of the anisotropy on the roughness of the surface as well. 
 
The Kondo resistivity of thin films was 
calculated \cite{UZ2,UZ3} 
assuming that the two surfaces of the thin 
films contribute to the anisotropy in an additive way, e.g.  
$K_{d,t}=\frac{\alpha}{d}+\frac{\alpha}{t-d}$, which was  
justified later in Ref.~\cite{Fomin}. 
The Kondo temperature was found only 
slightly affected \cite{UZ2} in a sample of finite size, 
in agreement with the experiments \cite{BG,exp1}.

\begin{figure} 
 \centerline{\epsfig{figure=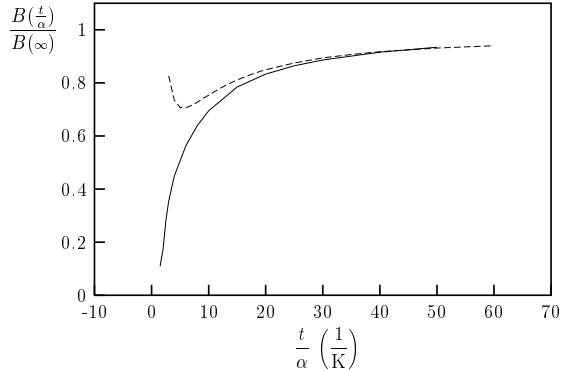,width=8.5truecm}} 
    \caption{Calculated size dependence in the Kondo resistivity 
      amplitude due to surface anisotropy. Solid line for $S=2$ and 
   $T_K=0.3$ K (i.e., $Au(Fe)$), dashed line for $S=5/2$ and 
   $T_K=10^{-3}$ K (i.e., $Cu(Mn)$). The minima in the half-integer case 
   may be only a sign of the breakdown of the weak coupling 
   calculation \cite{UZ3}.} 
\label{figBt} 
\end{figure}

The Kondo signal becomes reduced because close 
to the surface the motion of the Kondo spins is hindered by the  
spin-anisotropy \cite{UZ2,UZ3}, but 
the size dependence in the Kondo resistivity amplitude $B(t)$ defined by 
$\Delta\rho_{\rm Kondo}=-B(t)\ln T$ is different for integer and 
half-integer spins. For integer spins (e.g. $S=2$ for $Fe$)  
an impurity close enough to the surface is frozen to the  
${\bf n}{\bf S}=0$ state and the Kondo effect  
is impossible. Therefore  the  
amplitude $B(t)$ is reduced with respect to its bulk value  
and goes to zero as the film thickness is decreased (see 
Fig.~\ref{figBt}) \cite{UZ2}.  
For half integer spin (e.g. $S=5/2$ for $Mn$), on the other hand,  
the lowest energy state is the ${\bf n} {\bf S}=\pm 1/2$ doublet,  
thus the impurity 
still produces Kondo resistivity. As a consequence,  
the size dependence is much weaker, and $B(t)$ remains finite  
even for $t\rightarrow 0$ (see 
Fig.~\ref{figBt}) \cite{UZ3}. These results are
in agreement with the experiments \cite{BG,exp1}. For $S=1/2$ 
spin alloys (e.g. ${La}_{1-x} {Ce}_x$ films) no size  
dependence is expected due to the surface 
anisotropy in agreement with the experiment of Ref.~\cite{Roth}. 
\noindent 
\noindent  
The proximity effects \cite{prox,proxdis} can be well explained by the  
surface  
anisotropy as the number of available spin-orbit scatterers is increased  
by the overlayer and the magnetic impurities are in further distances
from the surface of the samples, but only if the overlayer is in the 
ballistic region
as well. In a new experiment of Giordano \cite{Giordanolayer} different  
multilayers composed of $Au$ and $Au(Fe)$ films were examined (where the  
overlayer was positioned only on one side, or on both sides of the film), 
giving good agreement also quantitatively with the predictions of 
the theory of surface anisotropy. 
 
There are experiments where quantities different from the Kondo resistivity 
were measured in order to test the theory of anisotropy as well. 
First Giordano measured the magnetoresistance \cite{Giordanomag} of thin films 
and found also a size dependence as the splittings due to the magnetic 
field and the surface anisotropy compete. 
Magnetoresistance calculations 
\cite{Borda} gave excellent agreement with these measurements. 
Thermopower \cite{Strunk} and  impurity spin magnetization measurements  
\cite{Hadok} on samples with reduced dimensions 
can also be explained by the theory of surface anisotropy.  
 
\subsection{Size dependence in point contacts} 
\label{S2B} 
Parallel to the thin film experiments, a thorough study of the  
Kondo effect in ultra small  $CuMn$ point contacts (PCs) has been  
carried out \cite{Yanson}. Rather surprisingly, in this case not a  
suppression but an {\em orders of magnitude increase} of   
both the Kondo signal and  the Kondo temperature has been reported.  
 
\begin{figure} 
\epsfxsize=7cm 
\epsfbox{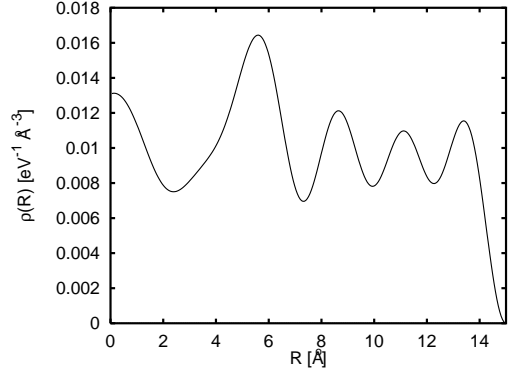} 
\caption{\label{fig:ldos}
Local density of states at the Fermi energy calculated for a 
point contact of diameter $d=30\AA$ and length $15\AA$.}
\end{figure}

\begin{figure} 
\epsfxsize=7cm 
\epsfbox{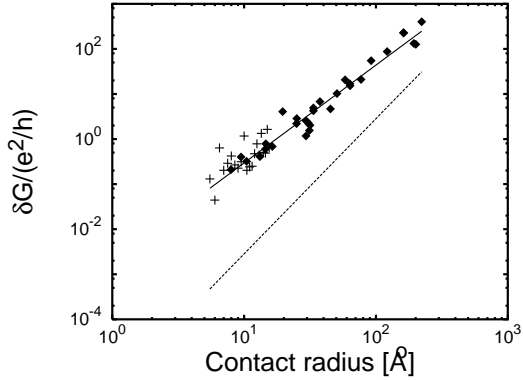} 
\caption{\label{fig:sizedep} 
Size dependence of the amplitude of the dimensionless Kondo 
 conductance $\Delta g=\Delta G\; h/e^2$ 
  for the model PC of Ref.~\protect{\cite{ZarandUdvardi}} as a
 function of the contact radius.  
Diamonds denote the experimental data 
 taken from Ref.~\protect{\cite{Yanson}} while theoretical  results  
(with no adjustable parameter) are indicated by crosses.  The 
 dashed line indicates the results without LDOS fluctuations ($g 
 \sim R^{3}$) 
  while the continuous line corresponds to the best fit to the 
  data of Ref.~\protect{\cite{Yanson}}: $g\sim 
R^{2.17}$.} 
\end{figure}

As  shown in Ref.~\cite{ZarandUdvardi}, these anomalies  can be  
well explained by the presence  of LDOS fluctuations: For a  
small PC, even a weak channel quantization induces huge  
LDOS fluctuations \cite{ZarandUdvardi} which become larger and larger  
with  decreasing contact sizes (see Fig. \ref{fig:ldos}). 
As $T_K$ depends on the LDOS  
exponentially (see Eq.~(\ref{eq:T_K})),  
this  may produce an extremely wide distribution of the Kondo  
temperatures for impurities in the contact region.  
The zero bias anomaly of the PC, however, turns out  
to be dominated by magnetic impurities with the largest $T_K$,  
since these are the ones that show a well-developed Kondo  
resonance. Indeed, in Ref. \cite{ZarandUdvardi}
the effect of these fluctuations was taken into account 
through a modified renormalization procedure, and 
a perfect agreement was found  
between the calculated and experimentally determined  
anomalous amplitude of the Kondo  signal 
(see Fig.~\ref{fig:sizedep}).

It was also predicted by the theory \cite{ZarandUdvardi}  
that this  effect should be much  
less pronounced for alloys with large $T_K$ as $T_K$ is less sensitive 
to the change in $\varrho_0$ in that case (see Eq.~(\ref{eq:T_K})),  
which has indeed been later confirmed by the experiments studying 
$Cu(Fe)$ alloys \cite{A}.

\subsection{Kondo resonance in the density of 
 states measured by STM} 
\label{S2C} 
 
It has been known for  a long time \cite{Mezei} that the local 
electron density of states nearby a magnetic Kondo impurity has a specific
structure due to the Kondo resonance. In the early experimental  
attempts a change in the electron density of 
states due to a layer of dilute magnetic alloys fabricated inside a 
metal has
 been measured \cite{Berman} by an oxide tunnel junction 
placed in a few atomic distances 
from that layer, and the 
Kondo structure was indeed observed. 
 
\begin{figure} 
\centerline{\epsfig{file=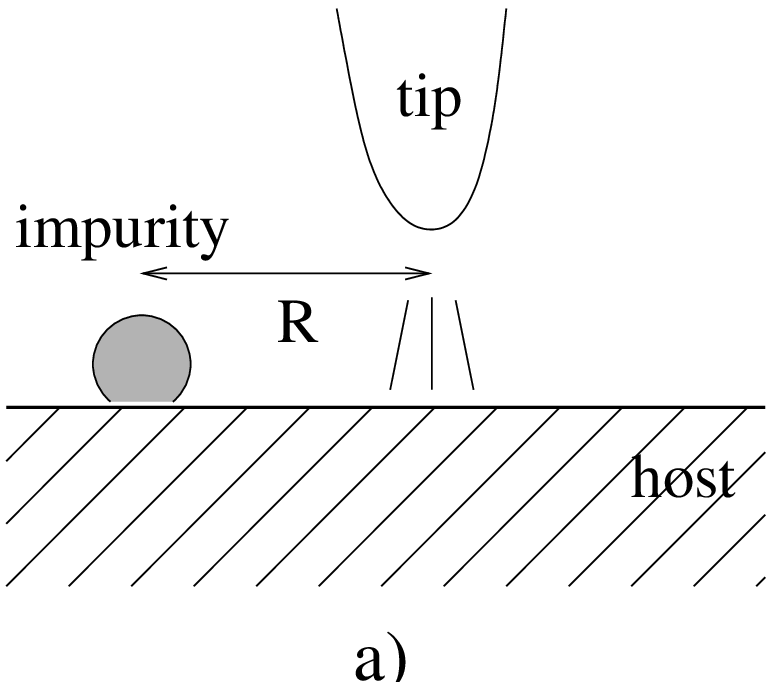,width=4.5truecm}} 
\vskip0.5truecm 
    \epsfig{file=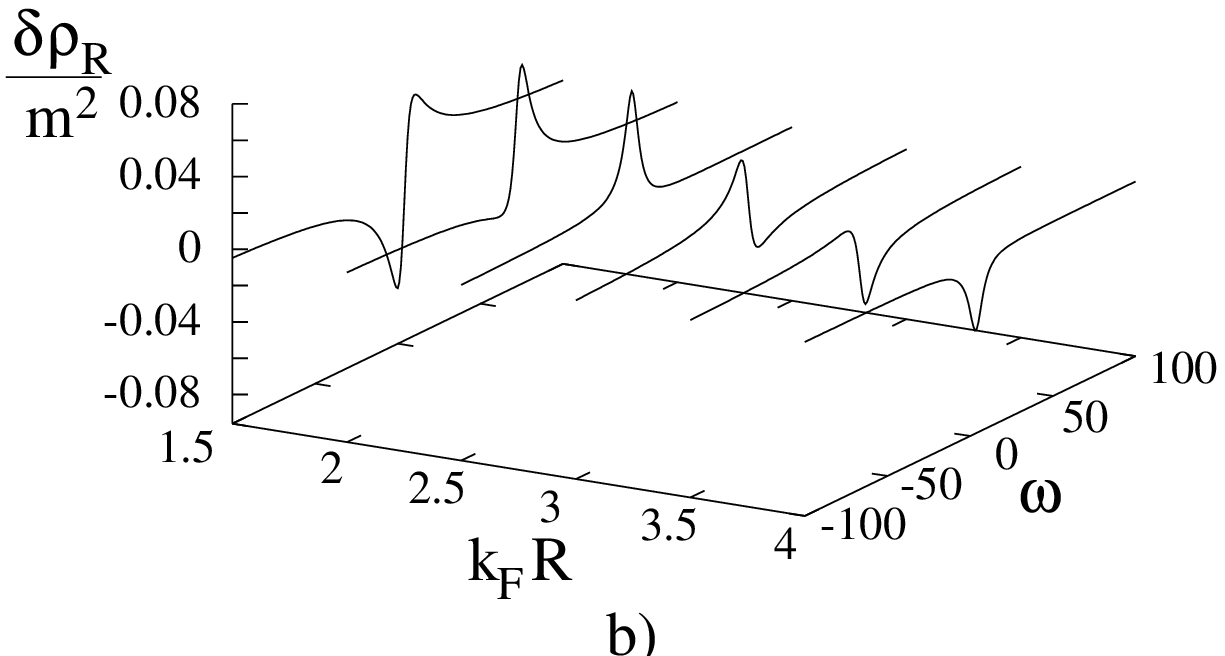,width=7.5truecm} 
    \caption{(a) Schematic plot of the measurement's setting. 
    (b) Qualitative dependence of the calculated line shape of the  
    tunneling DOS on the 
    distance of the tip from the impurity using the first part of  
    Eq.~(\ref{roveg}) and the 2-dimensional free-electron like 
    $Au(111)$ surface band Green's functions with effective mass 
    $m=0.26 m_e$ and Fermi energy $\varepsilon_F=0.52$eV \cite{Chen}. 
   The results obtained by using 3-dimensional bulk states are quite
    similar. \cite{UKSZZ}.} 
    \label{fig:1} 
\end{figure}

Recently, several groups have demonstrated using scanning tunneling 
microscopy (STM)  
\cite{Li,Madhavan,Manoharan} that a magnetic Kondo impurity 
adsorbed on the surface of a normal metal produces a narrow, resonance-like 
structure in the electronic surface density of states (DOS), whose asymmetric 
line shape resembles
that of a Fano resonance \cite{Fano}. 
The experiments were performed with single $Ce$ atoms on $Ag$  
\cite{Li} 
as well as with single $Co$ atoms on $Au$ \cite{Madhavan} and $Cu$ 
\cite{Manoharan} surfaces by measuring the I-V 
characteristics of the tunneling current through the tip of a 
STM placed close to the surface 
and at a small distance $R$ from the magnetic atom 
(see Fig.~\ref{fig:1} (a)). 
A  systematic study of the local electronic structure of individual 
transition-metal impurities on Au 
surfaces was performed by Jamneala {\em et al.}  
\cite{Jamneala} who showed that for elements near the 
end of the $3d$ row ($Ti$, $Co$, and $Ni$) the above mentioned 
narrow resonance structure appears, whereas for the elements around   
the center of the row ($V$, $Cr$, $Mn$, $Fe$) the electronic structure 
is found to be featureless. 
In these experiments the electron tunnels from 
the tip into the metal, travels to the impurity and, after 
scattering off it, goes back to the tip, resulting in an interference 
between the unperturbed and scattered electrons. There is also a 
possibility that the electrons tunnel from the tip directly to the 
magnetic impurity, more precisely into the d- or f-level of 
the atom. 
However, the tunneling rate for the latter process
is probably very small, especially for  f-levels, which are 
deeply inside the atom. 
 
The first theory proposed in Ref.~\cite{Madhavan,Schiller} takes into 
account both processes. Recently, it has been shown in 
Ref~\cite{UKSZZ}, however, that  
the Fano  resonance can develop even if one neglects the   
direct tunneling to the impurity. 
In this theory \cite{UKSZZ} the physics is governed by  
the unperturbed one-electron Green's function at the surface of the metal  
${\mathcal G}^{(0)}_{R,\sigma}(\omega-i\delta)$  
and the scattering amplitude  
$t_{\sigma}(\omega-i\delta)$ due to the impurity. 
The latter, given by $\frac{\Delta}{\pi\rho_0} G_{d,\sigma}(\omega-i\delta)$ 
in the Anderson model \cite{Anderson,UKSZZ}, can be approximated as 
\begin{eqnarray} 
  \label{eq:tomega} 
   &&t_{\sigma}(\omega-i\delta)= 
     \frac{\Delta}{\pi\rho_0}\biggl ( 
    \frac {Z_{d}}{\omega - \overline{\varepsilon}_{d} -i\Delta} 
    \nonumber \\ 
  &&+\frac {Z_{U}}{\omega -\overline {\varepsilon}_{d} -\overline{U} 
    - i\Delta} +\frac {Z_K}{\omega - {\varepsilon}_K - iT_K}\biggr)\ , 
\end{eqnarray} 
where $G_{d,\sigma}(\omega-i\delta)$ is the d-electron Green's 
function, $Z_d$, $Z_U$ and $Z_K$ are the appropriate strength of the 
poles, $\overline{\varepsilon}_{d}$, $\overline{U}$, $\Delta$, and  
$\varepsilon_K$ are the energies of the singly and doubly occupied 
orbitals of the effective model \cite{UKSZZ}, the broadening 
of the d-level, and the position of the Kondo resonance, respectively 
\cite{UKSZZ}. 
The final expression for the tunneling density of states reads 
\begin{equation} 
\label{roveg} 
  \delta\rho_{R}(\omega) = 
     \frac{[\Im {\mathcal G}^{(0)}_{R}(\omega^-)]^2} 
        {\pi\rho _0}\, 
       \biggl \{ 
  \frac{(q_{R}+\varepsilon)^2}{\varepsilon^2+1}- 1+ C_R 
       \biggr \} 
\end{equation} 
where the spin index $\sigma$ was dropped, $\omega^-=\omega-i\delta$, 
$\varepsilon=(\omega-\varepsilon_K)/{T_K}$  
and $q_{R}={\Re {\mathcal G}^{(0)}_{R}(\omega^-)} 
/{\Im {\mathcal G}^{(0)}_{R}(\omega^-)}$.
$C_R$, which depends on $Z_d$, $\overline{\varepsilon}_{d}$, $\Delta$, 
and $q_{R}$ \cite{UKSZZ}, arises from potential scattering on the 
d-level and corresponds to a weakly energy dependent 
Friedel oscillation.  
The first part of Eq.~(\ref{roveg}) coming 
from the scattering by the Kondo resonance gives a Fano line 
shape in the tunneling LDOS, controlled by the parameter $q_{R}$. 
The fit on the experimental data for a Co atom on a 
Au (111) surface \cite{Madhavan} gave excellent  
agreement with fitting parameters being consistent with the predictions 
of an NCA calculation combined  with band structure 
results \cite{UKSZZ}. 
 
In the experiment of Jamneala {\it et al.} \cite{Jamneala} the Kondo  
resonances were not observed in case of $V$, $Mn$, $Cr$, and $Fe$ 
atoms. In case of  
$Mn$, the Kondo temperature is small for bulk samples and it is further  
reduced by the weaker exchange coupling at the surface, thus the resonance 
cannot be expected on meV scale.
In case of $Fe$ and $Cr$ the surface  
anisotropy described by Eq.~(\ref{aniz}) \cite{UZ} may be also 
reduced, but even in that case, that may make impossible to see spin $S=2$. 
In case of $Co$ according to the  
electronic structure calculations \cite{UKSZZ}, the spin on the surface is  
close to $S=1/2$, where the anisotropy does not play a role.  
 
To calculate the distance dependence of $q_{R}$ and $C_{R}$, i.e., of the  
line shape, the tunneling of electrons from the tip (1) 
into the 3-dimensional $Au$ bulk states as well as (2) into the 2-dimensional 
$Au(111)$ surface band \cite{Chen} was considered. In both cases a free  
electron-like band structure was assumed \cite{UKSZZ}. 
Whereas the periodic changes of the line shape between Fano  
and Lorentzian ones and the decrease in the overall amplitude with  
increasing distance were demonstrated (see Fig.~\ref{fig:1} (b)), 
the precise dependence of 
the line shape on $R$ is not reproduced by our 
simplifying assumption of a free electron band structure \cite{UKSZZ}. 
That will require taking into account the 
detailed band structure as well as the additional scattering phase shift 
induced by the charge of the Co ion and the charge distribution around 
the $Co$ ion. 
 
Finally, it is worth to comment on the Fano parameter. In the original 
paper of Fano \cite{Fano} the parameter $q$ is defined in a way where it is 
proportional to 
$q^2\sim {|(\Phi|T|i)|^2}/{|(\Psi_E|T|i)|^2}$. 
The discrete level and the continuum state of energy $E$
have the wave function $\varphi$ and  
$\Psi_E$, respectively, and there is  hybridization between them with
amplitude $V_E$.
The initial electron state $|i\rangle$ has a transition  
described by operator $T$ to the exact states including the hybridization. 
The discrete state $\Phi$ in definition of $q^2$ is, however, not the 
original unhybridized state 
$\varphi$, but the state modified by the hybridization as 
$\Phi=\varphi+{\mathcal P}\int dE' {V_{E'}\Psi_{E'}} (E-E')^{-1}$ \cite{Fano}. 
Thus $(\Phi|T|i)$ can be different from zero even if $(\varphi|T|i)=0$,  
i.e., thus even without 
direct transition to the localized state $\varphi$, 
a transition rate still exists into the state $\Phi$. 
That clearly shows that Fano line shape can be obtained 
without direct transition to the localized d- or f-states in the present case. 
 
\section{Possibility of two-channel Kondo effect due to structural 
  defects} 
\label{S3} 
 
\begin{figure} 
\centerline{\epsfig{figure=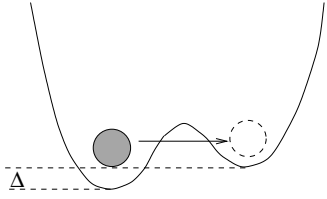,width=4.3truecm}} 
    \caption{The potential for a tunneling atom between two positions 
    (TLS) is shown, where the arrow represents the tunneling and the 
    asymmetry splitting is $\Delta$.} 
\label{figtls} 
\end{figure}

It is well established by now that scattering on fast dynamical  
defects can produce Kondo-like anomalies\cite{Kondo,VladZaw}.  
In the simplest model  the defect atom tunnels between  two  
positions and thus forms a  two-level system (TLS).  
These two levels are typically  split due to the  spontaneous  
tunneling between the positions  
and the asymmetry of them, resulting in a typical splitting  
of $\Delta \sim 1-100K$ (see Fig.~\ref{figtls}). 
\noindent 
\noindent  
In the TLS Kondo model  
the coordinate of the dynamical impurity is coupled to the angular  
momentum of the conduction electrons through an effective exchange  
interaction, and the real spins of the conduction electrons  
act as silent channel indices. Consequently --  
in the absence of splitting -- 
the physics of the TLS is described by the two-channel Kondo model  
predicting a NFL behavior below $T_K$. In this model the spin-flip 
scattering of the original Kondo model is replaced by electron 
assisted tunneling.   
 
\subsection{Point contacts} 
\label{S3B} 
 
\begin{figure} 
\epsfig{figure=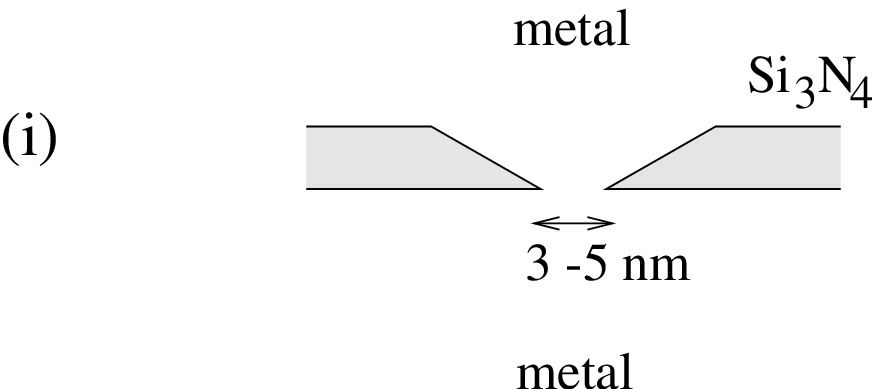,width=6truecm} 
\vskip0.3truecm 
\epsfig{figure=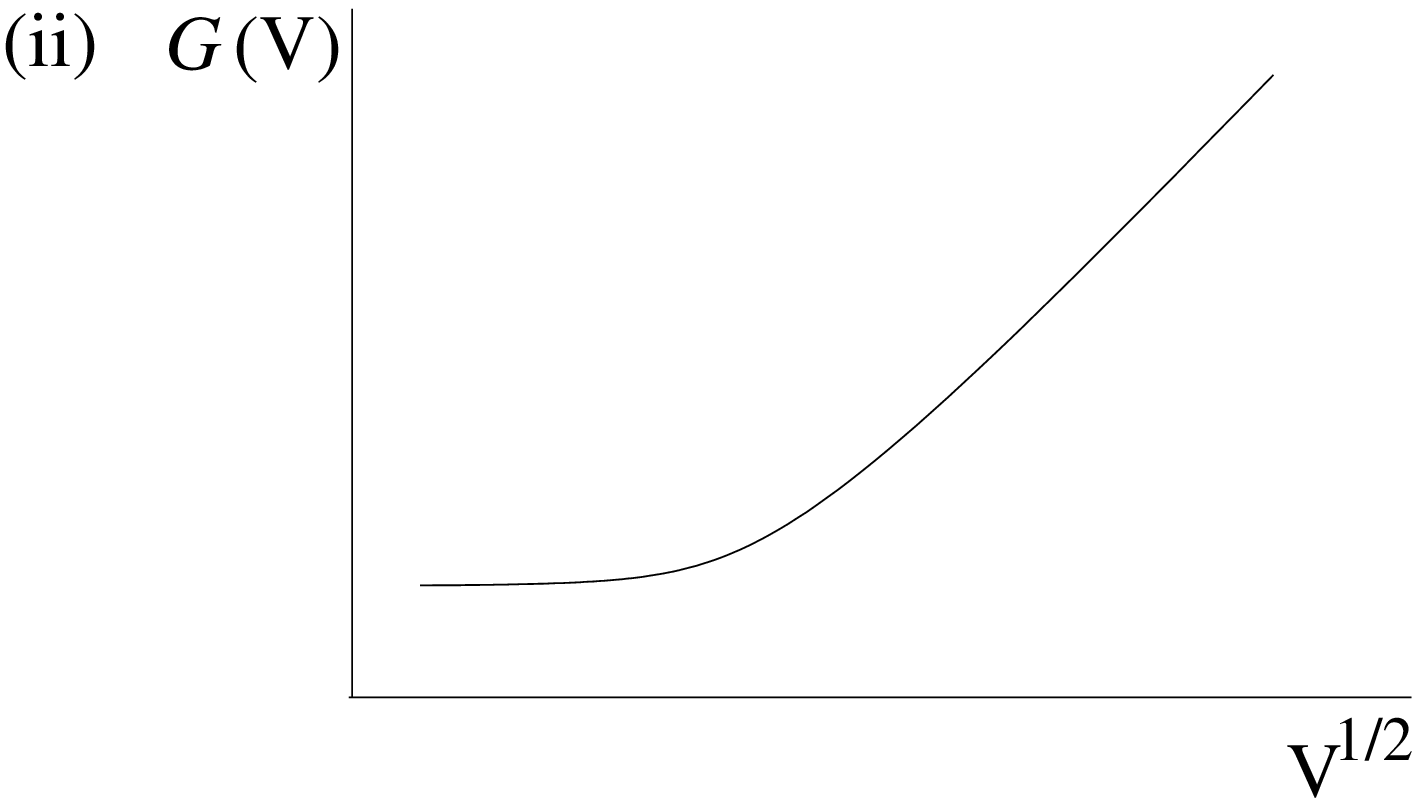,height=4.1truecm} 
    \caption{(i) Point contacts used in Ref.~\cite{RalphBuhrman,Upad}. 
(ii) Schematic plot of the observed low voltage dynamical conductance 
${\it G}(V)$ exhibiting a NFL like behavior $V^{1/2}$ which shows a 
crossover to FL like behavior at very low voltage which is either due 
to the temperature \cite{RalphBuhrman} or other low energy cutoff like 
splitting $\Delta$ \cite{Upad}.} 
\label{figpc} 
\end{figure}

Several experiments have been reported where the observed 
low temperature  anomalies were attributed to TLS 
Kondo defects \cite{PbGeTe,disloc,RalphBuhrman,Upad,Keijsers,Zarcomm} 
(see Fig.~\ref{figpc}). 
\noindent 
\noindent  
In all these experiments the observed 
anomalies were unambiguously due to {\em dynamical} structural 
defects: they disappear under annealing and  not or only 
slightly depended on magnetic field. 
 
A logarithmic increase of the resistivity 
attributed to the presence of dislocations or substitutional 
tunneling impurities has been 
observed in various systems \cite{PbGeTe,disloc}. 
However, the most spectacular experiments were carried out in $Cu$ and 
$Ti$ point  contacts where a two-channel Kondo-like  $\sim\sqrt{T}$ 
and $\sqrt{V}$ non-Fermi liquid scaling behavior  
due to non-magnetic scatterers has been observed in the contact resistance 
\cite{RalphBuhrman,Upad}. The widths of the zero bias anomalies
were associated with the Kondo temperature, $T_K \sim 5K$. 
In another beautiful experiment a fluctuation of the zero bias anomaly 
between two curves due to some slow TLS's has been observed  in
amorphous point  
contacts \cite{Keijsers}, which could be consistently explained  
assuming that a slow fluctuator influences the splitting of  
one or two fast Kondo two-level systems close to it \cite{Zarcomm}.  
There is a further experiment \cite{balkashin} where an alternating 
voltage was superimposed on a constant bias $V_0$, 
$V(t) = V_0 + V_1 \cos(\omega t)$. As far as the characteristic 
frequency (e.g. Kondo temperature) of the mechanism responsible for the 
zero bias anomaly is large compared to $\hbar \omega$ the 
measured $I-V$ characteristic is just the time average of 
the current: $\langle{I(t)}\rangle = I (V_0) + {1\over 4} \left({\partial^2 I\over 
\partial V^2}\right)_{V = V_0} V_1^2$.  
 For frequencies higher than this scale an $1/\omega$ dependence  is expected.
No deviations have been  observed experimentally 
even for $\nu = 60$ GHz ($2.4$ K) implying 
that $T_K> 5 {\rm K}$. This lower bound is in agreement with the value of $T_K$
estimated from the width of the zero-bias anomaly.

There remain, however, a number of puzzles. In all these 
experiments the estimated Kondo temperature is 
in the range $T_K \sim 10 K$. $T_K$ has been first estimated  
in  Ref.~\cite{VladZaw} assuming 
that  TLS's are formed by a {\em heavy atom} that tunnels  
within a distance of about $\sim 0.4\AA$, and was found  to be in the  
range $\sim0.01-1K$. It has been suggested that virtual hopping to the  
lowest excited states could increase $T_K$ substantially \cite{ZarZawTLS},  
however, in the above model this  turned out to be wrong  
\cite{Altshexc,Zarconf}. It  
has been shown that,  in reality,  $T_K$ is reduced even further  
if one includes the effect of  {\em all} the excited states.  
Thus, within the original simplistic 
TLS model, where the TLS is formed by  
some heavy  atom tunneling between two close positions and electron-hole 
symmetry is assumed, it seems to be  
impossible  to have  $T_K$ in the experimentally observed range.  
It has been pointed out recently, that the criticism of 
Ref.~\cite{Altshexc} is essentially based on 
the assumption of electron-hole symmetry in the 
conduction electron density of states \cite{zaw_zar_comm}. 
Electron-hole symmetry, however, is strongly violated in  any 
realistic band  structure. As shown in Ref.~\cite{zaw_zar_comm}, 
$T_K$ can be  enhanced by orders of magnitude
with a relatively small electron-hole symmetry breaking even when  all 
excited states are included, and $T_K$ can be in a much broader range 
than expected  previously.

On the other hand, one has very little knowledge about the  
microstructure of the TLS's, and in order to make any quantitative  
prediction it would be extremely important  
to identify it. We expect that   TLS's with a relatively {\em small 
effective mass} (such as Hydrogen stuck at the surface of the sample  
 or  dislocation jogs \cite{Sethna}) could be  able to tunnel over a 
distance of 
$\sim 1\AA$ and probably produce a  $T_K$ in the experimentally observed 
range. 
 
Another interesting question is related to the splitting of the two levels, 
which provides a lower cutoff for the NFL scaling. 
The presence of splitting and the cutoff of NFL behavior 
has been observed in several experiments. 
In particular, measurements on  $Ti$ point contacts are in perfect 
agreement with all  predictions of the TLS Kondo model. 
In Ref.~\cite{RalphBuhrman}, however, 
the number of TLS's has been estimated to be about 50, 
for which concentration already  
a significant deviation from the NFL scaling  
should have appeared due to the presence of disorder generated
splitting \cite{Smolyarenko}. 
However, no such deviation has been reported in Ref.~\cite{RalphBuhrman}. 
The resolution for this puzzle may  
also be related to the precise microstructure of the tunneling  
impurities. 
 
\subsection{Electron dephasing time $\tau_{\phi}$} 
\label{S3C}

\begin{figure} 
\centerline{\epsfig{figure=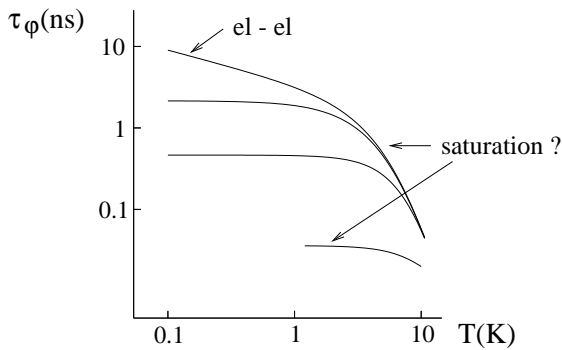,width=7.5truecm}} 
    \caption{Schematic plot of the temperature dependence of the 
    dephasing time, $\tau_{\varphi}$ for cases where the el-el 
    interaction dominates and where there are strong deviations from 
    them (saturation ?).} 
\label{figsat} 
\end{figure}

Recent developments in mesoscopic physics  
raised an interesting 
question about  the electronic dephasing time  
$\tau_{\phi}$.  This is the time scale for an electron  
to stay in a given exact one-electron state  in the presence  
of static impurities. The transitions between these states are 
due to electron-phonon, electron-electron, electron-dynamical defects (e.g. 
 two-level system), or electron - magnetic impurity interactions.  
At low temperature the electron-phonon interaction freezes out and the 
 electron-electron interaction becomes dominant which has been studied by 
 Altshuler and his  collaborators \cite{altshuler}. 
 According to that theory the dephasing time tends to infinity as the 
 temperature is lowered, as the available phase space for the 
 electron-electron scattering gradually vanishes. 
However, as recently  emphasized 
by Mohanty and Webb \cite{mohanty}, experimentally this is not always 
 the case. In some materials and samples, like $Ag$ produced 
 by the Saclay group \cite{saclay}, $\tau_{\phi}$ follows very closely the 
 predictions of the electron-electron interaction theory 
but in other cases there are strong deviations from the predicted 
behavior and the data indicate some saturations (see 
Fig.~\ref{figsat}).  
\noindent 
\noindent  
It has been known since 
a long time that the saturation-like behavior depends very much on the 
preparation of the samples \cite{label4} and even on the substrates on which 
the films are deposited \cite{label5}. The early suggestion by Lin and 
Giordano \cite{label4} was that the dephasing is either due to the magnetic 
impurities or some defects which are very sensitive to the metallurgical 
properties of the films including thickness, annealing etc.  
The effect of magnetic Kondo 
impurities on the dephasing rate was carefully studied in those cases where 
the Kondo temperatures were in the relevant temperature range \cite{label6}.  
There the dephasing rate shows a maximum at the Kondo  
temperature of the magnetic impurities due to the enhancement of  
spin-flip  scattering, but at 
lower temperature  singlet Kondo ground state is 
formed and $\tau_\phi$ decreases as the spin-flip rate  gradually  
freezes out (see Table~\ref{table}). However, the saturation observed
has no resemblance to this behavior.  
 
The only possibility for magnetic  impurities to produce a  
saturation of $\tau_\phi$ would be if their Kondo temperatures  
were much smaller than the temperature range of interest. Then  
the spin-flip rate is approximately temperature independent.  
However, in order to have $T_K\leq 10 mK$ the exchange coupling must  
be very weak $J\rho_0\leq 3\times 10^{-2}$ and to produce the  
dephasing rate observed an enormous number of unidentified magnetic  
impurities should be present, which is very unlikely. 
 
Accepting that the low temperature dephasing anomalies are  
intrinsic properties of the samples and far from a universal  
behavior it looks reasonable that some 
local dynamical defects as TLS's are responsible for them. 
 
Depending on the electron-TLS interaction two different limits  
must be considered: (i) For  weak couplings the electron induced  
transition in the TLS is treated  in  second order perturbation  
theory. In that case to get an almost temperature independent dephasing  
rate the splittings $\Delta$ (excitation energies) must be smaller than  
the measured temperature and their distribution must be peaked  
at very low energies \cite{label8}. However, there is no evidence  
for such anomalous distribution, and linear specific heat measurements 
on metallic glasses are consistent rather with  a uniform 
 distribution \cite{label7}.  
(ii) The other theoretical possibility is given  
by TLS's with 2CK behavior \cite{label9}.  
In that case, in contrary to the magnetic Kondo problem, the 
scattering rate at low temperature is due to processes 
where the final states contain many electron-hole pairs (see
Table~\ref{table}), and being a 
dynamical scattering process, this produces  dephasing.
\begin{table*}
  \begin{center}
    \epsfig{figure=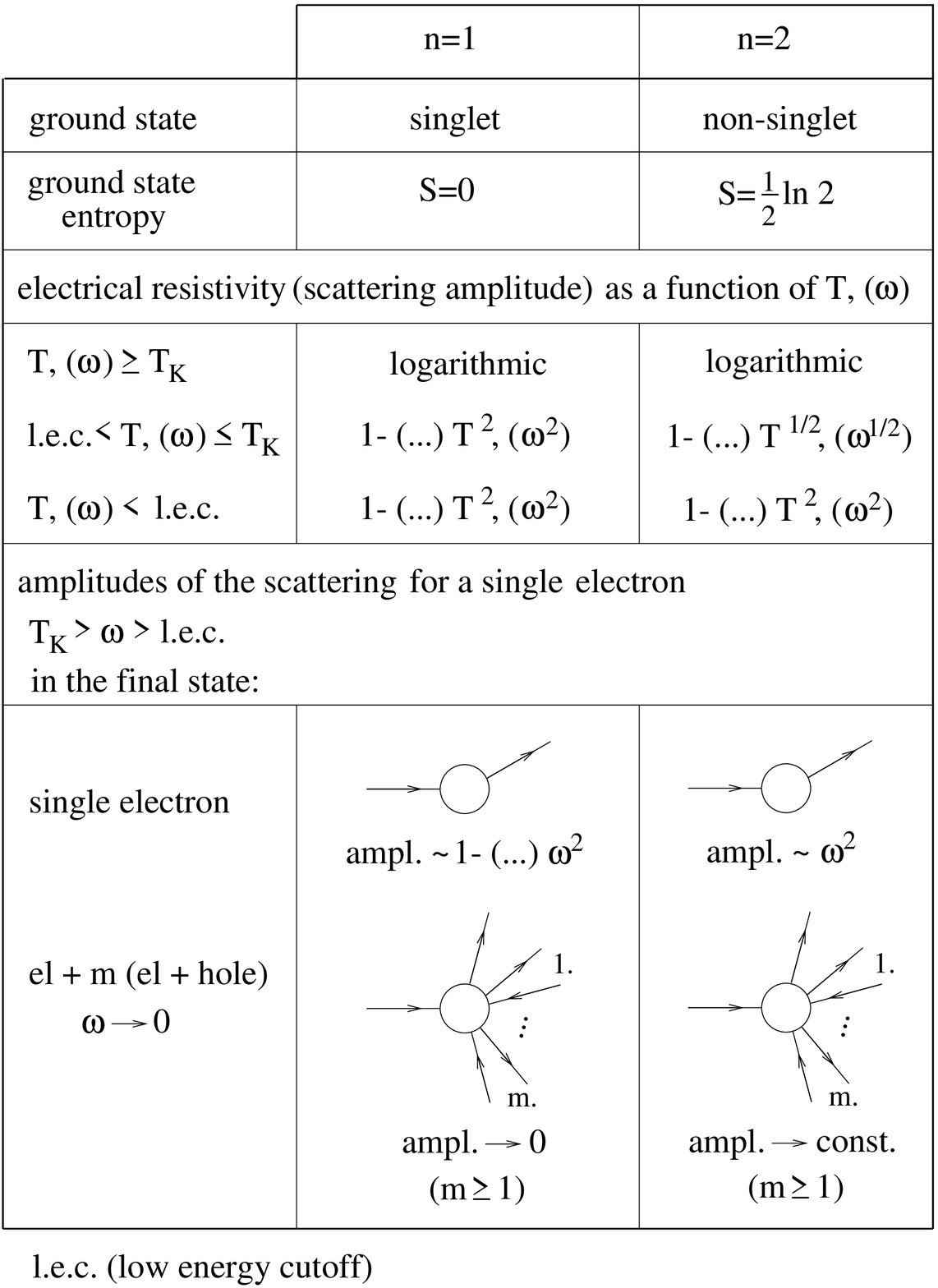, height=14cm}
  \end{center}
  \caption{Comparison of the $n=1$ and $n=2$ channel Kondo problems.}
\label{table}
\end{table*}  
In order to get a reasonable dephasing rate less than $1 ppm$ 2CK  
defect is required. This  explanation has two drawbacks: 
The questionable existence of such 2CK defects and 
the required small splitting $\Delta$, even if this latter is
 renormalized 
downwards due to the strong interaction by a factor $\Delta/T_K$ if 
$\Delta < T_K$ \cite{label10}. On the other hand in case of 2CK  
defects the non-universality  and metallurgical dependence  
are quite natural. 
 
Finally it should be mentioned that such saturation like behavior has also  
been seen in degenerate semiconductor ballistic quantum dots  
\cite{label11}. In degenerate semiconductor as far as we know even the 
theory of magnetic Kondo effect has not been worked out in detail. 
 
\subsection{Energy distribution of electrons in short wires} 
\label{S3D} 
 
\begin{figure} 
\epsfig{figure=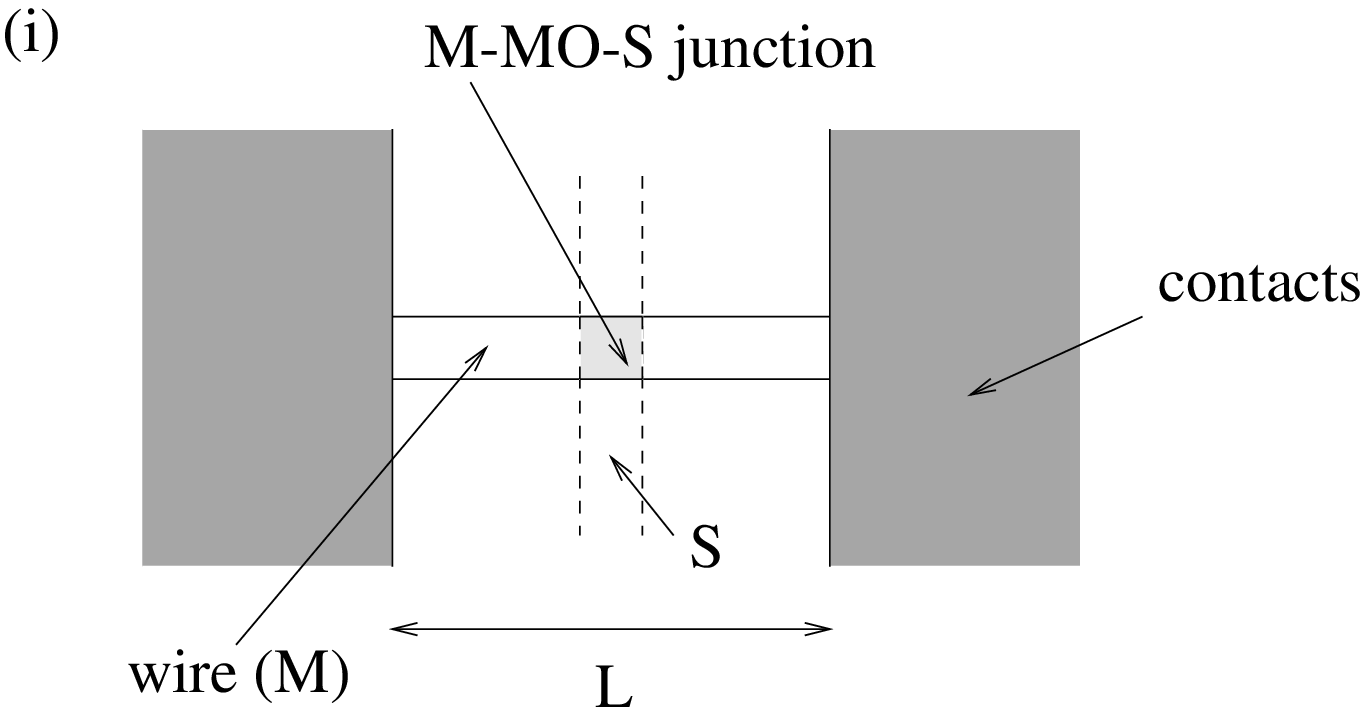,width=7.5truecm} 
\vskip0.4truecm 
\epsfig{figure=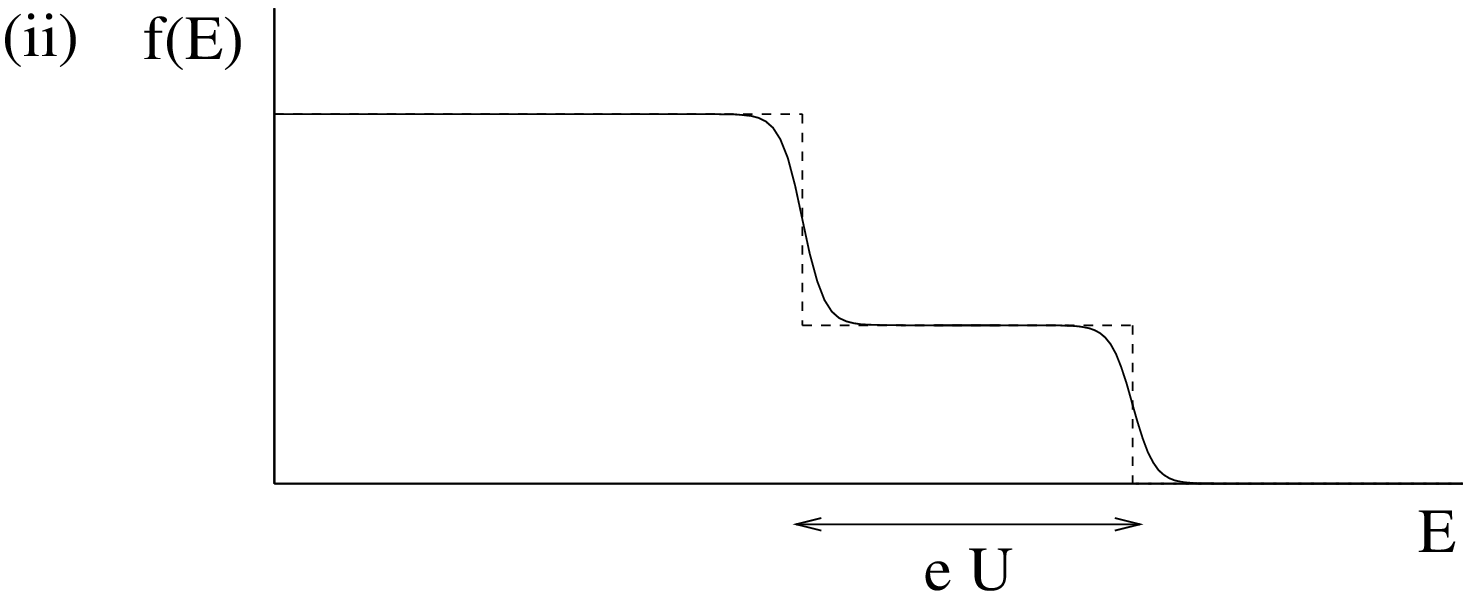,height=3truecm} 
    \caption{Saclay experiment: (i) layout of the experiment with 
    metallic wire and the measuring M-MO-S junction. (ii) Electron 
    distribution function nearby the middle of the 
    wire. Non-interacting case (dotted line) with two steps separated 
    by the applied voltage $U$ and the schematic measured curve (solid 
    line) with smeared steps.} 
\label{figmmos} 
\end{figure}

In addition to the dephasing problem another closely related dilemma exists 
concerning the energy distribution of electrons in short wires with finite 
bias voltage. In these experiments \cite{label12,label13}  
one measures the energy distribution function of electrons 
$f(E)$ by fabricating a metal-metal oxide-superconductor (M-MO-S) 
tunnel junction  
at various positions along the wire (see Fig.~\ref{figmmos}). 
\noindent 
\noindent  
From the $I-V$ characteristic $f(E)$ is  determined by deconvolution.  
The typical length of the wires was $1.5 \mu m$ and $5 \mu m$.  
The measurements were carried out at 
$25 mK$, and the samples were in the diffusive limit. 
In this case the typical energy relaxation processes are slow  
compared to the time it takes an electron to diffuse  
through  the sample. Therefore  the electron distribution  
exhibits typically two steps corresponding to 
the Fermi energy of the left and right contacts,  
thus their difference is proportional to the applied bias  
$U$ \cite{label14}. For non-interacting electrons at distance  
$x$ measured from one of the contacts the distribution 
function is 
\begin{equation} 
 f_x(E,U)=\left(1-{x\over L}\right)f^0(E+eU)+{x\over L}f^0(E), 
\end{equation} 
where $L$ is the length of the sample and $f^0$ is the  
equilibrium distribution. In the diffusive limit without  
energy relaxation these two steps are smeared only 
by the temperature. At low enough temperature this smearing  
becomes, however, much larger than $T$, and the measured smearing gives 
 information on the relaxation processes.  
In the case of long samples the 
 smearing can be essential as the electrons spend longer time in the sample 
 before leaving into the contacts. Typically the  applied voltage  
$U\sim 0.1-0.2 meV$ is larger  than the temperature  $T\sim30 mK$.  
The surprising results for short $Cu$ wires were that the shape and  
the amplitude of the smearing could not be explained by the  
electron-electron interaction. Using a 
Boltzmann equation approach the line shape could only be  reproduced by  
assuming an anomalously strong electron-electron interaction kernel $K$  
with  an anomalous dependence on the 
transferred energy  $K\sim1/\varepsilon^2$ \cite{label12,label13}, 
clearly in disagreement  
with  the predictions of the  theory of electron-electron  
interaction \cite{altshuler}. On the other hand, it has been 
shown many years ago \cite{solyom_zawa} that such dependence 
can be due to magnetic impurity mediated inelastic scattering 
for $T>T_K$. The same dependence is valid for two-channel Kondo 
impurities. For $Ag$ samples, however, a good  
agreement has been found with the electron-electron interaction theory 
 with the expected energy dependence $K\sim\varepsilon^{-3/2}$  
and amplitude. 
 
The close similarity of this problem to the dephasing time 
dilemma became obvious, when the dephasing time  
was directly measured on samples prepared in the same way,  
and it was found that $\tau_\phi$  in the $Ag$ 
samples \cite{label15} follows again the standard electron-electron 
interaction theory and does not saturate. 
 
$Au$ wires prepared in Saclay, on the other hand, show also a  strong 
 relaxation rate and exhibit an electron distribution that  could be  
fitted using a $K\sim\varepsilon^{-2}$ kernel.  
The strong anomalies in  the electron distribution  
and the dephasing time  of $Cu$ and $Au$ wires  
and the fact that  $Ag$ wires behave ``regularly'' 
both in dephasing time and electron distribution experiments 
strongly indicate that the anomalous dephasing time and  
 distribution function are related to  the same non-universal  
and material dependent processes. 
 
It has been realized \cite{KrohaZawadowski}  
that assuming a  $K\sim\varepsilon^{-2}$ electron-electron  
interaction kernel for energies $E\gg kT$  
an  energy independent electron relaxation rate can be derived  
making a connection between the distribution function  
 dephasing time experiments. It has been shown 
that the $Au$ and $Cu$ experiments can be well explained  
by  2CK impurities  with negligible splitting \cite{KrohaZawadowski} 
by treating the 2CK problem in the framework of non-crossing 
 approximation (NCA) \cite{label17} and handling the non-equilibrium  
situation within the quantum Boltzmann equation framework.  
 
That method had been applied earlier to the non-equilibrium transport 
through nano-point contacts in the presence of 2CK defects \cite{label18}. 
Kroha also showed that the electron-electron  
interaction mediated by 2CK centers shows a 
$K\sim\varepsilon^{-2}$ energy dependence \cite{label19}.   
That suggests  that the classical Boltzmann equation with that kernel  
may give similar 
 results to those obtained by using quantum Boltzmann equation and the 2CK 
 scattering. 
 
 The most challenging feature of the experiments is that for an 
 intermediate range of applied bias $U$, $0.1 mV < U < 0.5 mV$ the measured 
 distribution function follows a scaling 
 \begin{equation} 
 f_x(E,U)=f_x({E\over U}). 
 \end{equation} 
 
 In the framework of the 2CK interpretation the space-dependent 
 non-equilibrium quantum Boltzmann equation is solved, where the
 collision term 
 is expressed by the non-equilibrium electron self-energy   calculated in 
 a self-consistent way. In these calculations $T_K > 1 K$ is assumed 
 which determines the energy scale of the 2CK effect with zero applied bias. 
 The Kondo effect is due to the sharp step of the unperturbed energy 
 distribution function at the Fermi edge and the size of the step determines 
 the Kondo temperature. As far as the bias $U$ is not larger than  $T_K$,  
from the point of view of the Kondo effect there are no two separate 
 steps and no scaling holds as the presence of $U$ influences the 
 shape, width and amplitude of the Kondo resonance.  For  $T_K <  U$,  
however the two steps are separated,  and  two distinct Kondo 
 resonances are formed  at the energy of each step 
with an effective Kondo temperature which can be essentially 
 smaller than the equilibrium one. That bias region can be described
 by scaling according to Kroha's  NCA results \cite{label19}. At even higher 
 voltage other parameters of the model start to be essential and the 
 scaling breaks down again. 
 This  theory is in good agreement with the experimental results for 
 the  complete applied 
 voltage range $U$ and different positions of the measuring M-MO-S diodes and 
 for samples of various length. For $Cu$ and $Au$ samples the
 anomalous smearing can 
 be obtained but for $Au$ a much higher concentration of 2CK centers is needed 
 ($100 ppm$) and for $Cu$ $1-5 ppm$. These values are in complete accordance 
 with the amplitude of the measured dephasing rate. It should be 
 emphasized that all  
 these data, taken from the measurements of the $Cu$, $Au$ samples prepared 
 in Saclay, very likely depend a lot on the 
 sample preparation, metallurgy and maybe also on the substrates used. 
 
\section{Discussions and perspectives} 
 
The spin Kondo problem in mesoscopic systems has been thoroughly 
studied both experimentally and theoretically and is quite well 
understood. There remain, however, a few further questions to answer:
Concerning the surface anisotropy, 
the crossover from the ballistic to the  dirty limits and 
the effect of disorder on the surface anisotropy should be further 
clarified. Similarly to the surface anisotropy, local density of 
states fluctuations decay  as $1/d$ as a function of the distance 
from the surface. These fluctuations give probably the dominant 
effect in very thin films and films with 
a weak spin-orbit interaction for alloys with a relatively 
small $T_K$. Measurements on a host with weak 
spin-orbit scattering could help to clarify these issues.

The observation of the Kondo resonance by STM due to 
a single Co atom on the  surface is a very impressive technical achievement. 
In the future, it would be worthwhile to study magnetic 
impurities inside the first few surface layers to establish stronger 
coupling between the spin and the host metals. In order to understand 
the data or to make predictions 
further electronic calculations are 
required for the host metal at the surface, the charge redistribution 
due to the impurity and the value of spin at the impurity atom. 
 
The anomalous behavior of the zero bias anomalies in 
point contacts, and the dephasing and the transport in short wires can be 
well described by dynamical impurities with 2CK behavior. The fact 
that not all
the samples studied (like $Ag$ prepared in Saclay) 
show the anomaly supports that the anomalous NFL behavior is 
intrinsic, non-uniform and very likely depends on the preparation and 
treatment of the samples, and even on the substrate on which the sample 
is deposited. 
The orbital 2CK interpretation of these experiments 
is very challenging, however, 
it is far from being well established. 

The possibility of the 2CK effect 
from a heavy tunneling atom has been suggested longtime ago. 
The original version of this model, however, cannot 
explain the large Kondo temperature $T_K\approx 1\sim 10$K.
However, as realized recently, electron-hole symmetry breaking 
present in all realistic band structure based density of states,
can increase $T_K$ substantially and is very promising 
to resolve the long-standing problem of $T_K$.
That is certainly the most important question to answer
concerning the application of  the idea of the 2CK problem. 

Finally, we want to remark that a more detailed literature of this
topic will be found in the Proceedings of the NATO 
workshop on 'Size dependent magnetic scattering' held between 28 May
and 1 July 2000 in P\'ecs, Hungary \cite{NATO}.
 
\section{Acknowledgements} 
 
We thank all our collaborators and colleagues
for fruitful discussions, 
especially the participants of the NATO 
workshop held between 28 May and 3 July 2000 in P\'ecs and Budapest, 
Hungary. A. Z. benefited from the hospitality of the Meissner
Institute and LMU in Munich where he was supported by the Humboldt 
Foundation. Our research was supported by the Hungarian Grants OTKA
T029813, T024005, T030240, and F29236.

\end{document}